\documentclass[useAMS,usenatbib]{mnras}
\usepackage{graphicx}
\usepackage{natbib}
\usepackage{amssymb}
\usepackage{setspace}
\usepackage{epsfig,color}
\usepackage{multicol}
\usepackage{xcolor}
\usepackage{float}
\usepackage{hyperref}
\usepackage{upgreek} 
\usepackage{longtable}

\def\lint{\int\limits} 
\def\D{\mathrm{d}}

\newcommand{\Msol}{~$M_{\odot}$} 	% solar mass
\newcommand{\cc}{~cm$^{-3}$}		% cm^-3
\newcommand{\gsqc}{~g\,cm$^{-2}$}	% g/cm^2
\newcommand{\gcc}{~g\,cm$^{-3}$}	% g/cm^3
\newcommand{\rhopdf}{$\rho$-PDF}  
\newcommand{\Npdf}{$N$-PDF}  
\newcommand{\bPLFIT}{{\sc bPlfit}}

\title[Multiple power-law tails in PDFs]{Multiple power-law tails in the density and column-density distribution in contracting star-forming clumps}

\author[Veltchev et al.]
{	
\parbox{\textwidth}{Todor V.~Veltchev$^{1,2}$\thanks{E-mail: eirene@phys.uni-sofia.bg}, Philipp Girichidis$^2$, Lyubov Marinkova$^3$, Sava Donkov$^4$, Orlin Stanchev$^1$ and  Ralf S. Klessen$^{2,5}$}\vspace{0.4cm} \\
\parbox{\textwidth}{
  $^1$University of Sofia, Faculty of Physics, 5 James Bourchier Blvd., 1164 Sofia, Bulgaria\\
  $^2$Universit\"at Heidelberg, Zentrum f\"ur Astronomie, Institut f\"ur Theoretische Astrophysik, Albert-Ueberle-Str. 2, 69120 Heidelberg, Germany\\
  $^3$Department of Applied Physics, Technical University, 8 Kliment Ohridski Blvd., 1000 Sofia, Bulgaria \\
  $^4$Institute of Astronomy and NAO, Bulgarian Academy of Sciences, 72 Tsarigradsko Chausee Blvd., 1784 Sofia, Bulgaria \\
  $^5$Universit\"{a}t Heidelberg, Interdisziplin\"{a}res Zentrum f\"{u}r Wissenschaftliches Rechnen, Im Neuenheimer Feld 205, D-69120 Heidelberg, Germany
}
}

\date{Submitted 2024 January 4}

\pagerange{\pageref{firstpage}--\pageref{lastpage}} \pubyear{2024}
\begin{document}
\label{firstpage}
\maketitle

\begin{abstract}
We present a numerical study of the evolution of power-law tails (PLTs) in the (column-)density distributions ($N$-PDF, $\rho$-PDF) in contracting star-forming clumps in primordial gas, without and with some initial rotational and/or turbulent support. In all considered runs multiple PLTs emerge shortly after the formation of the first protostar. The first PLT (PLT~1) in the \rhopdf{} is a stable feature with slope $q_1\simeq -1.3$ which corresponds -- under the condition of preserved spherical symmetry -- to the outer envelope of the protostellar object with density profile $\rho\propto l^{-2}$ in the classical Larson-Penston collapse model, where $l$ is the radius. The second PLT (PLT~2) in the \rhopdf{} is stable in the pure-infall runs but fluctuates significantly in the runs with initial support against gravity as dozens of protostars form and their mutual tidal forces change the density structure. Its mean slope, $\langle q_2\rangle\simeq -2$, corresponds to a density profile of $\rho\propto l^{-3/2}$ which describes a core in free fall in the classical Larson-Penston collapse model or an attractor solution at scales with dominating protostellar gravity. PLT~1 and PLT~2 in the \Npdf{}s are generally consistent with the observational data of Galactic low-mass star-forming regions from {\it Herschel} data. In the runs with initial support against gravity a third PLT (PLT~3) in the \rhopdf{}s appears simultaneously with or after the emergence of PLT~2. It is very shallow, with mean slope of $\langle q_3\rangle\simeq -1$, and is associated with the formation of thin protostellar accretion disks. 
\end{abstract}


\section{Introduction}
\label{Introduction}
A large variety of investigation tools have been proposed to study the physics and the evolution of molecular clouds in their variety of star-forming activity. Widely used in the last two decades is the analysis of the probability density function (PDF) of mass density (\rhopdf; easily constructed from numerical data) and of column density (\Npdf). The popularity of this approach rests on many studies which have shown that the PDF shape and its distinctive parts bear imprints of the physical conditions and processes in the clouds. Supersonic turbulent flows in isothermal, non-gravitating medium lead to a lognormal  \rhopdf{} \citep{VS_94, Padoan_Nordlund_Jones_97, Passot_VS_98, Klessen_00, Li_Klessen_MacLow_03, Kritsuk_ea_07, FKS_08, Molina_ea_12}, with some possible deviations from lognormality in the low-density range \citep{Federrath_ea_10, Konstandin_ea_12, Pan_ea_19}. A more complex, bimodal \rhopdf{} with two peaks might result from thermal instabilities in purely hydrodynamic flows \citep{VS_ea_00}. In the high-density range, a power-law tail (PLT) of the \rhopdf{} can appear in case of non-isothermal compressive turbulence, with polytropic index less than unity \citep{Scalo_ea_98, Passot_VS_98}. But most typically, the development of high-density PLT out of a quasi-lognormal \rhopdf{} reflects the increasing role of self-gravity in evolving star-forming clouds \citep{Klessen_00, VS_ea_08, Kritsuk_Norman_Wagner_11, Collins_ea_12, Federrath_Klessen_12, Federrath_Klessen_13, Burkhart_ea_17}. The emerging PLT is steep and hardly distinguishable from a lognormal wing; later, in the course of local collapses, its slope becomes shallower and tends toward a constant value whereas its span increases toward lower densities \citep{Veltchev_ea_19, Khullar_ea_21}. This evolution was studied and substantiated theoretically by \citet{Girichidis_ea_14}, from analytical model of a collapsing homogeneous sphere, and by \citet{Jaupart_Chabrier_20} who showed that the PLT density range corresponds to regions in which gravity is the dominant agent in the statistics of turbulence.

The derivation of \Npdf{} from observations of star-forming regions is not straightforward. Often, the column density is calculated from the measured visual extinction or thermal dust emission; this involves some assumptions about dust properties and constraints on the column-density range and introduces uncertainties which are difficult to quantify. Nevertheless, observational \Npdf{}s can be appropriately corrected for effects of foreground contamination, observational noise and incompleteness \citep{Ossenkopf_ea_16}. They display a similar functional form like the \rhopdf{}s predicted from numerical and theoretical studies: a largely lognormal main part and a single PLT at the high-density end \citep[e.g.,][]{Kainulainen_ea_09, Froebrich_Rowles_10, Pokhrel_ea_16, Schneider_ea_13, Schneider_ea_15b} whose slope appears to correlate with the current stage of the cloud in the star formation process \citep{Abreu-Vicente_ea_15}. This overall picture is in agreement with the interpretation that, in general, the PLT phenomenon indicates the dominance of self-gravity during collapse. Indeed, the \Npdf{} evolution as investigated in simulations of contracting clouds is characterized by the emergence and further growth of a PLT \citep{BP_ea_11, Kritsuk_Norman_Wagner_11, Federrath_Klessen_13, Auddy_Basu_Kudoh_18, Koertgen_ea_19, Veltchev_ea_19}. If spherical symmetry and isothermality are acceptable approximations for the collapsing region which accounts for the PLT, the slope of the latter is interrelated to the density profile \citep{Donkov_Veltchev_Klessen_17}. 

Recently, there is a growing interest to the existence of {\it double} PLTs of (column-)density distributions in evolved star-forming clouds. First indications of a second PLT of \rhopdf{}s emerging at very high densities have been found by \citet{Kritsuk_Norman_Wagner_11} from purely hydrodynamic simulations of supersonic, isothermal and self-gravitating turbulent medium with grid refinement reaching AU scales in the dense cores. A small single PLT (PLT~1) was detected at the formation time of first collapsing objects; after $\sim0.4$ free-fall times it spans already 6 orders of magnitude  and a second, shallower PLT (PLT~2) grows out of its high-density end. The authors attributed this PLT~2 to support through centrifugal forces within rotating contracting cores. Further numerical magneto-hydrodynamic research of supersonic and self-gravitating turbulence confirmed the development of a PLT~2 of \rhopdf{} after the first PLT had been established \citep{Collins_ea_11, Collins_ea_12}. High-resolution simulations of turbulent collapse by \citet{Murray_ea_17} yield a first PLT with slope corresponding to an attractor solution for the density profile at small scales and a much shallower PLT~2 which is always associated with close vicinities of the forming protostars, within the typical radius of their disks. Thorough analysis of the double-PLT phenomenon in \rhopdf{}s of simulated isothermal gravoturbulent fluids was performed by \citet{Khullar_ea_21}. Those authors conclude that the density at which the first PLT deviates from the lognormal main part accounts for the transition from unbound to bound gas while the deviation point of the PLT~2 delineates structures in rotationally flattened disks.

In addition, significant theoretical efforts were made to explain the PLT~2. \citet{Donkov_Stefanov_19} model an ensemble of self-gravitating, isothermal turbulent clouds with small homogeneous core, power-law \rhopdf{} and steady-state accretion. They obtain two different slopes in the regimes far from the core and near to the core; the latter is steeper and indicative of free fall. In a follow-up study \citep{Donkov_ea_21}, their model was modified through a different treatment of gas thermodynamics in the regime far (isothermality) and near (an equation of state of a ``hard polytrope'') to the cloud core; in the second case, a polytropic exponent of $4/3$ can account for a shallower PLT~2. In contrast to these two works, \citet{Jaupart_Chabrier_20} consider the \rhopdf{} as a dynamical entity and develop an analytical theory to describe its evolution in supersonic star-forming cloud as the balance between gravity and turbulence changes. In their framework, the PLT emerges at the time of starting collapse in some regions while a shallower PLT~2 forming at later stages is associated with dense clumps in free-fall collapse. 

On the part of observations, there is a growing evidence for double PLTs in \Npdf{}s in star-forming regions. The first reported cases were three high-mass regions whose \Npdf{}s display combination of a PLT and a shallower PLT~2 \citep{Schneider_ea_15a}. A more recent study from {\it Herschel} imaging of dozens Galactic regions with various mass and star-forming activity shows that the column-density distribution in most of them has a double PLT, with a shallower or a steeper PLT~2, while the main part is best fitted with two lognormals \citep{Schneider_ea_22}. The double PLT phenomenon at the typical spatial scales of {\it Herschel} observations is still poorly understood. In clouds with a low mass fraction of dense cores, PLT~1 cannot be attributed (only) to local collapses but rather to gravitational collapse and accretion at larger scales as indicated by a number of observational studies \citep[see references in][Sect. 4.2]{Schneider_ea_22}. Some suggested explanations of a flatter PLT~2 are feedback effects from young stars, such as ionization compression \citep{Tremblin_ea_14} or strong magnetic fields in initially subcritical clouds \citep{Auddy_Basu_Kudoh_18} while a steeper PLT~2 might result from the orientation of magnetic field with respect to the (column-) density distribution \citep{Soler_19}. 

The presented paper aims to contribute to the current research of the double PLT phenomenon and its suggested interpretation as signature of rotational support at the scales of protostellar disks. We study the evolution of PLTs both in \rhopdf{} and \Npdf{} from high-resolution simulations of contracting star-forming clumps with varied initial balance of energies which allows to distinguish the effect of rotation on the high-density end of the PDF. The applied technique for multiple PLT extractions does not depend on fitting of the main PDF part, in view of the uncertainty regarding its shape. Details on the used numerical data are provided in Section \ref{Data}. Section \ref{Method} informs the reader of the method to assess the parameters of PLTs. The results on the PLT evolution are presented in Section \ref{Results}; their implications, in comparison to other works, are discussed in Section \ref{Discussion}. Section \ref{Summary} gives a brief summary of this work. 

\section{Numerical data}
\label{Data}

We use data from a large set of numerical simulations of protostar formation in primordial gas by \citet[][hereafter, W20]{Wollenberg_ea_20}. They have been performed with the Voronoi moving-mesh code {\sc Arepo} \citep{Springel_10}, adapted with a number of modifications (for details, see W20 and the references therein). The code enables us to adjust the grid resolution according to the growth of density fluctuations and thus is very suitable to investigate collapse in star-forming regions \citep[e.g.,][]{Greif_ea_11, Greif_ea_12} and, in particular, the evolution of high-density parts of the (column-)density distribution with a good enough resolution.

The W20 simulations follow the gas collapse in a box with a sidelength of $13$~pc which initially contains a single Bonnor-Ebert (BE) sphere at its centre, with the respective density profile \citep{Bonnor_56, Ebert_55}. For simplicity, we label this object hereafter as {\it contracting star-forming clump}. The linear radius of the sphere is $R_{\rm BE}=1.87$~pc, corresponding to the critical dimensionless value for stability, and the mass within it is $M_{\rm BE}=2671$~\Msol. Since W20 have explored the formation of Population III stars, the gas metallicity is zero and the initial temperature inside the sphere was set to 200~K which yields its Jeans content: $M_{\rm BE}/M_{\rm J}\simeq 3$, where $M_\mathrm{J}$ is the critical mass for the stability of self-gravitating isothermal spheres \citep{Jeans_1902}.  A network of 45 chemical reactions between different species of H and He and free electrons is used to compute cooling and the polytropic index $\gamma$ (see W20, Sect. 2.1). The numerical box outside the BE sphere is filled with homogeneously distributed gas of density $\rho(R_{\rm BE})=2.7\times10^{-21}$\gcc. With this setup, the impact of multi-scale flows and externally driven accretion onto the central clump are neglected.

We stress that we are primarily interested in the global collapse problem of a gravitationally unstable structure rather than in the details of the primordial conditions in terms of metallicity, chemical reactions and the resulting cooling\footnote{The low metallicity and cooling rate might have some effect on clump fragmentation which takes place at scales below 1000~AU.}. We further note that the clouds in W20 are strongly dominated by gravitational contraction. Consequently, it might be expected that the main features of this process do not differ substantially from those in collapsing clumps in the present-day star-forming regions. Hence, a comparison of the W20 simulations with observational data in the local Universe is instructive and appropriate. 

Different physical setups for the BE sphere were chosen through varying the initial amounts of rotational and/or turbulent energy: 
\begin{itemize}
 \item Pure infall (PI) motions: no turbulence and no rotation.
 \item Rotation only (RO), no turbulence: we choose between a low ($0.01$) or high ($0.10$) value of the ratio $\beta$ between rotational kinetic and gravitational potential energy. In these cases, the rotation around the $z$-axis is accounted for by setting a uniform angular velocity. 
 \item Turbulence only (TO), no rotation: here, we select a  low ($0.05$) or high ($0.25$) value of the ratio $\alpha$ between turbulent and gravitational potential energy.
 \item Rotation and turbulence (RT): these are simulations  with  all possible combinations of the values of $\alpha$ and $\beta$ specified above.
\end{itemize}

Five runs (realizations) have been performed for each setup; those without turbulence differed only in the random mesh positions for initialization of the BE density distribution, whereas those with turbulence differed in their random seeds. To avoid artificial fragmentation, a Jeans refinement criterion is applied so that the local Jeans length is always resolved with at least 16 cells. The formation of protostars in the collapsing gas was traced through sink particles. The latter were defined on several standard conditions (see W20, Sect. 2.2); for the goals of the present study it is important to highlight the threshold (minimal) density for sink formation $n_{\rm th}=2.4\times10^{15}$\cc{} ($\rho_{\rm th}\sim 4\times10^{-9}$\gcc) and the accretion radius $r_{\rm acc}=2$~AU which controls the collapse and the gravitational boundedness of gas in the close vicinity of a sink particle. In fact, $r_{\rm acc}$ defines the effective spatial resolution achieved in the densest parts of the collapsing region, and $\rho_{\rm th}$ constitutes the upper density limit for our investigation of PLTs in the \rhopdf. 

For our study, we select a sample of eight W20 runs: two simulations of type PI, four of type RO,  as well as one TO and one RT realization. Their main parameters are given in Table~\ref{table_W20_runs}. The point in time, at which the first sink particle forms is labeled `star formation time', $t_{\rm SF}$  (Column 3). A single sink particle forms and grows in mass by accretion from the surrounding envelope material in the PI runs, whereas in all other ones there is significant fragmentation, leading to the formation of small clusters of accreting objects (Column 5). The simulation runs up to $t=t_{\rm end}\sim 10^3$~yr after $t_{\rm SF}$ (Column 4), when the ionizing feedback from the formed stars can no longer be neglected. At that time the accretion rate onto sink particles effectively drops to zero (cf. Fig. 2 in W20) and the star-formation efficiency (Column 7) approaches a constant value of order of a few percent.

\begin{table}
\caption{Runs from the W20 simulations selected to study the PLT evolution. Notation: PI -- pure infall, $\alpha$ -- the ratio between turbulent and gravitational potential energy, $\beta$ -- the ratio between rotational and gravitational potential energy, $t_{\rm SF}$ -- the time of formation of the first sink particle, $\tau_{\rm ff}\simeq0.34$~Myr -- free-fall time of the simulated cloud, $t_{\rm end}$ -- run duration after $t=t_{\rm SF}$, $N_{\rm sink}$ -- number of sink particles at $t_{\rm end}$ and their total mass, $M_{\rm sink}$.}
\label{table_W20_runs} 
\begin{center}
\begin{tabular}{lc@{~~}c@{~}c@{~~~}c@{~~}cc}
\hline 
\hline 
Realization & $\beta$ &  $t_{\rm SF}$ & $t_{\rm end}$ & $N_{\rm sink}$ & \multicolumn{2}{c}{$M_{\rm sink}$}  \\ 
~    & ~ & [ $\tau_{\rm ff}$ ] & [ yr ] & ~ & [\Msol{}\,] & [\,$M_{\rm BE}$\,] \\
\hline 
PI $-$ 3 & -- & $1.86$ & $~784$ &	$~1$ & $53.2$ & $2.0\,\%$ \\
PI $-$ 5 & -- & $1.86$ & $~786$ &	$~1$ & $55.4$ & $2.1\,\%$ \\
$\beta001-2$ & 0.01 & $1.89$ & $1048$ &	$26$ & $51.6$ & $1.9\,\%$ \\
$\beta001-4$ & 0.01 & $1.89$ & $1179$ &	$43$ & $55.9$ & $2.1\,\%$ \\
$\beta010-2$ & 0.10 & $2.19$ & $1677$ &	$31$ & $38.3$ & $1.4\,\%$ \\
$\beta010-4$ & 0.10 & $2.19$ & $1889$ &	$27$ & $36.4$ & $1.4\,\%$\\
$\alpha025-1$ & -- & $2.05$ & $1910$ &	$72$ & $67.9$ & $2.5\,\%$\\

$\alpha025\beta010-1$ & 0.10 & $2.34$ & $1087$ &	$41$ & $53.4$ & $2.0\,\%$\\
\hline 
\hline 
\end{tabular} 
\end{center}
\smallskip 
\end{table}

\section{Method for PLT extractions}
\label{Method}

For the analysis of the high-density parts of $\rho$-/\Npdf s from the selected W20 runs, we apply a technique which allows for extraction of multiple PLTs. It is an extension \citep{Marinkova_ea_21} of the so-called {\it adapted \bPLFIT{} method} for PLT extraction \citep{Veltchev_ea_19}. Below we summarize the most important features of this technique. Items (i) to (iv) concern the adapted \bPLFIT{} method in general and item (v) describes the used technique to extract multiple PLTs. 
\begin{enumerate}
 \item The adapted \bPLFIT{} method is based on the statistical approach \bPLFIT{} \citep{Virkar_Clauset_14} aimed to assess the {\it assumed} power-law part\footnote{The existence of a better fit through some non-power-law function is not ruled out.} of an arbitrary binned distribution via Kolmogorov-Smirnov goodness-of-fit statistics. In contrast to other commonly used techniques, the PLT is extracted without any assumptions about the possible functional shape (e.g., lognormal) of the rest of the distribution. This is an advantage in comparison with other approaches, especially when the main part of the PDF suffers from incompleteness and/or has a complex shape, probably affected by multiple competing physical agents. 
 \item The input for the method consists of three parameters: (a) the lower density cutoff controls the range to search for a PLT; (b) the upper density cutoff excludes the poorly resolved high-density tail of the PDF (e.g., the threshold for sink formation $\rho_{\rm th}$ in analysis of \rhopdf s); (c) range of variation of the total number of bins $N_{\rm bins}$ within the chosen cutoffs, depending on the data volume and on the span of the expected PLT: e.g., $30\le N_{\rm bins}\le 100$.
 
 \item The output of the method for {\it fixed} $N_{\rm bins}$ are the two parameters of the suggested PLT: deviation point (DP) and slope. This output is not sensitive to the chosen binning scheme: linear, logarithmic, etc. In general, it is also weakly affected by the choice of $N_{\rm bins}$ / bin size -- as demonstrated via applications of \bPLFIT{} to analytical PDFs composed of a lognormal function and a PLT \citep[][Appendix B]{Veltchev_ea_19}. Nevertheless, working with PDFs from discrete data sets, some choices of $N_{\rm bins}$ may lead to extraction of unreliable PLTs as the method fits a few bins at the end of the distribution. Such false PLTs are excluded from further consideration by setting some minimal span of the extracted PLT in units of bin size (see Fig. 2 in \citealt{Veltchev_ea_19}).
 \item The obtained set of reliable PLTs (after exclusion of false PLTs) is used for {\it averaging the PLT parameters} over $N_{\rm bins}$ within the range of variation. 
\item The extraction of multiple PLTs proceeds in the following main steps (Sect. 4.2 in \citealt{Marinkova_ea_21}): (a) The lower cutoff is gradually increased, starting from the lower data limit, and averaged PLT parameters are obtained for each lower cutoff choice (output sample). The upper density cutoff is set at $\rho_{\rm th}$; (b) The number of PLTs and their approximate parameters are made by eye; points with high standard deviation from these initial estimates are excluded from the output sample (see Fig. 3 in \citealt{Marinkova_ea_21}); (c) The DP of a PLT is set as the upper density cutoff of the previous PLT (if there are indications of its existence) and step A is re-done for the latter; (d) The data from the final output sample are used to average the PLT parameters over the groups associated with the presumed PLTs.
\end{enumerate}

 In principle, the technique can distinguish between PLTs with very similar slopes. Therefore additional criteria for plausible extractions of {\it multiple} PLTs might be imposed depending on the PDF shape, the data resolution and/or the expected values. For instance, one can require a {\it minimal difference in slopes} between two extracted PLTs and a {\it minimal density span} of each one, in order to consider both as `true detections'. In \citet{Marinkova_ea_21}, these control parameters were set to $0.4$ and one order of magnitude, respectively. 
 
 In this work, we do not impose any preliminary criteria for plausible PLT extractions, in view of the dynamics and complexity of the PDF shape and of our goal to trace the emergence and the development of a PLT in small time steps in regard to the dynamical time. Additional inspection of the PLT parameters was performed once the results for the whole considered run of W20 had been obtained. Then, at time points when {\it both} the differences in slope and DP of two successive PLTs turn out to be less than the uncertainties of these PLT parameters, the PLTs were merged into a single one. For examples, see the evolution of \rhopdf{}s for the runs with $\beta=0.10$ in Fig. \ref{fig_PLT_evolution_RO_density}, right.

\section{Evolution of the PLTs}
\label{Results}

In this Section we present the emergence and the evolution of multiple PLTs in the course of the selected W20 runs. Logarithmic density $s=\ln(\rho/\rho_0)$ and logarithmic column density $z=\ln(N/N_0)$ are defined as normalization units using the mean density of the initial Bonnor-Ebert sphere $\rho_0\equiv\langle\rho\rangle_{\rm BE}=3M_{\rm BE}/(4\uppi R_{\rm BE}^3)=6.6\times 10^{-21}$\gcc~and its corresponding column density $N_0\equiv\rho_0 \times 13~{\rm pc}=0.26$\gsqc. Multiple (up to three) PLTs of the \rhopdf{} and the \Npdf{} are given by the equations:
\begin{equation}
 p(s)=\left\{ \begin{array}{ll}
              A_{s, 1}\exp(q_1 s),& s_1 \le s < s_2 \\
              A_{s, 2}\exp(q_2 s),& s_2 \le s < s_3 \\
              A_{s, 3}\exp(q_3 s),& s \ge s_3 
             \end{array} \right.
\end{equation}
and
\begin{equation}
 p(z)=\left\{ \begin{array}{ll}
              A_{z, 1}\exp(n_1 z),& z_1 \le z < z_2 \\
              A_{z, 2}\exp(n_2 z),& z_2 \le z < z_3 \\
              A_{z, 3}\exp(n_3 z),& z \ge z_3~~, 
             \end{array} \right.
\end{equation}
where $(q_i, s_i)$ and $(n_i, z_i)$ are the corresponding two PLT parameters (slope and DP) and $A_{s, i}$ and $A_{z, i}$ are constants, for $1\le i\le3$. We point out that the upper (column-)density ends of PLT~1 and PLT~2 parts do not necessarily coincide with the DPs of the next PLT (see the comment in Sect. \ref{rho-PDFs}).

Since the development of PLTs is expected to take place in the course of local collapse(s) we apply our technique to the evolutionary times after the formation of the first sink particle to the end of the run: $t_{\rm SF}\le t \le t_{\rm end}$. At each time step, after the initial detection of all PLTs through the adapted \bPLFIT, their PLT parameters were re-evaluated, setting the DP of the high-order PLT as upper density cutoff $s_{\rm max}$ -- e.g., taking $s_3\equiv s_{\rm max}$ to obtain new estimates of $(q_2, s_2)$. This is a small modification in comparison to \citet{Marinkova_ea_21}, to avoid possible effects of higher-order PLTs on the parameters of the lower-order ones.

\begin{figure} 
% % \hspace{3.5em}
\begin{center}
\includegraphics[width=84mm]{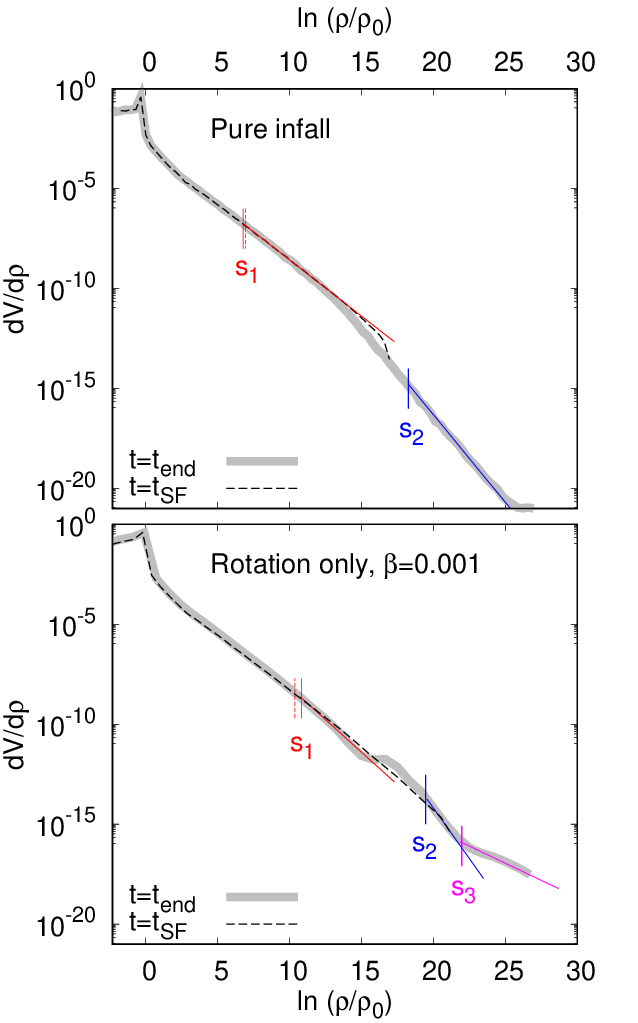}\\
\caption{Density PDFs at $t=t_{\rm SF}$ (dashed) and $t=t_{\rm end}$ (thick grey) in case of pure infall (top) and initial weak rotation support (bottom). Only densities below the sink particle threshold $\ln(\rho_{\rm th}/\rho_0)=27.17$ are considered. The DPs are marked with the corresponding line notation; the slopes at $t=t_{\rm end}$ are plotted.}
\label{fig_rho_PDF_examples}
\end{center}
\end{figure}

\subsection{Density PDFs}
\label{rho-PDFs}

Examples of \rhopdf{}s (for some fixed $N_{\rm bins}$) at the onset of star formation and the end of the simulation are plotted in Fig. \ref{fig_rho_PDF_examples}. The main, non-power-law part of the density distribution deviates substantially from lognormal due to the setup of the simulated clumps, immersed into an initially homogeneous medium (Sect. \ref{Data}). A single PLT~1 spanning several orders of magnitude  is already present at the time of first sink particle formation $t_{\rm SF}$, both in case of pure infall and initial weak rotational support ($\beta=0.001$). At the end of the simulation, there is clear evidence of multiple PLTs. The DP $s_1$ of PLT~1 is approximately the same (cf. the dashed and solid red marks). A steeper PLT~2 is formed and spans more than 3 decades in density in the pure-infall runs (Fig. \ref{fig_rho_PDF_examples}, top). In addition, in the runs with $\beta=0.001$, a third, much shallower PLT (PLT~3) has grown out of PLT~2 at $t=t_{\rm end}$, reducing significantly its span (Fig. \ref{fig_rho_PDF_examples}, bottom). 

\begin{figure}
% \hspace{3.5em}
\begin{center}
 \includegraphics[width=83mm]{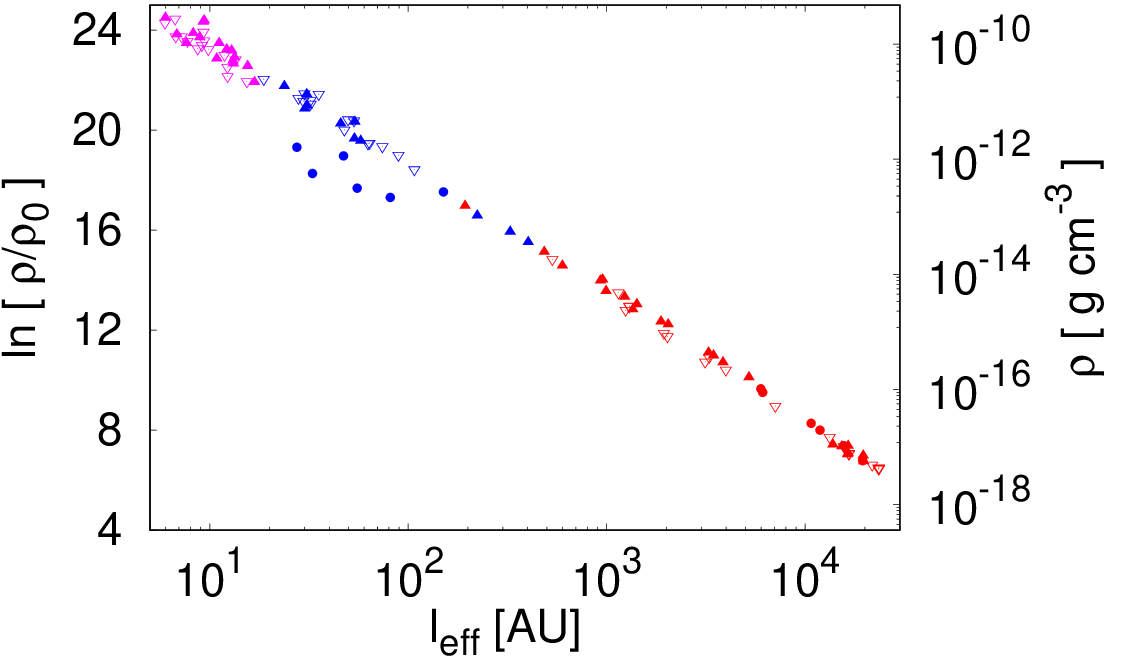}
\caption{Spatial scales corresponding to all extracted PLT 1 (red), PLT 2 (blue) and PLT 3 (magenta) from the runs with pure-infall (filled circles), $\beta=0.01$ (open triangles) and $\beta=0.10$ (filled triangles).}
 \label{fig_Size-slope_relation}
\end{center}
\end{figure}

It should be noted that our technique for multiple PLT detection tends to extract shorter PLTs in two cases: i) if the PDF contains large bump(s) in the considered density range\footnote{Such features appear at the late stages of runs with initial rotational and/or turbulent support as dozens of protostars form and their mutual tidal forces cause redistribution of material in their vicinities.}, and/or ii) there is a density range of smooth transition between two PLTs. For instance, after visual inspection of both panels in Fig. \ref{fig_rho_PDF_examples}, one might put by eye $s_2$ at some lower value and estimate $q_2$ accordingly. However, such a fitting function would produce a larger Kolmogorov-Smirnov statistics than the PLT estimate from \bPLFIT{} does and hence would be rejected by the method (see Sect. \ref{Method} in this work and Sect. 3 in \citealt{Marinkova_ea_21}). The result would be the same as $s_{\rm min}>s_1$ is being varied -- the adapted \bPLFIT{} method tends to extract a PLT with parameters which depend weakly on the $s_{\rm min}$ value.

\begin{figure} 
% % \hspace{3.5em}
\begin{center}
\includegraphics[width=75mm]{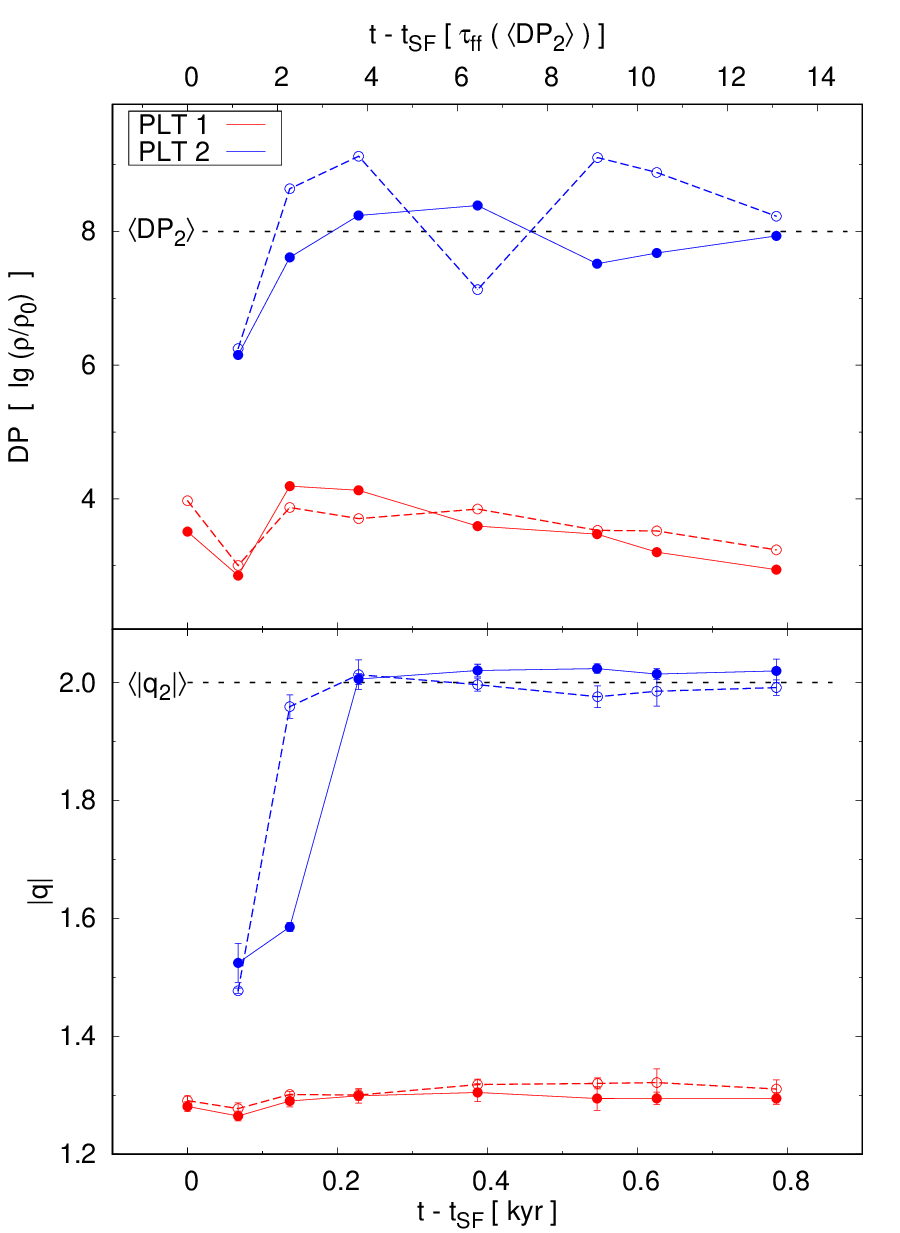}\\
\caption{Evolution of PLTs of the \rhopdf{} in realizations PI$-$3 (solid, filled symbols) and PI$-$5 (dashed, open symbols) of the pure-infall case, in terms of slope (bottom) and DP (top). Time on the second $x$-axis is given in units of free-fall time calculated for the mean density and slope within the PLT~2 range (black, long dashed; see text).}
\label{fig_PLT_evolution_PI_density}
\end{center}
\end{figure}

Fig. \ref{fig_Size-slope_relation} displays the spatial scales which correspond to the extracted PLTs from the studied runs with pure-infall motions as well those with some initial rotational support. The effective sizes are calculated as $l_{\rm eff}=(V(s_i)/V_{\rm tot})^{1/3}L$ where $L=13$~pc is the sidelength of the numerical box and $V(s_i)/V_{\rm tot}$ is the fraction of the total volume defined by the isocontours corresponding to the DPs $s_i$ ($1\le i \le3$).

\subsubsection{Pure infall}
\label{Pure-infall_rho_pdfs}

\begin{figure} 
\begin{center}
\includegraphics[width=83mm]{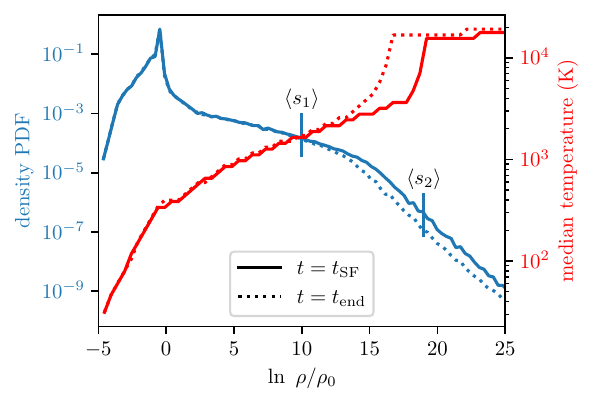}\\
\caption{PDF of the density and the median temperature as a function of density from a pure-infall run. The averaged DPs of PLT 1 and PLT 2 are shown. The change in slope in the density PDF coincides with a strong increase in gas temperature.}
\label{fig_PI_density_temperature}
\end{center}
\end{figure}

The evolution of the double PLT in the pure-infall runs is shown in more details in Fig. \ref{fig_PLT_evolution_PI_density}. It is very similar for the two studied realizations. PLT~1 has a slope $q_1\simeq-1.3$ at $t=t_{\rm SF}$ which remains unchanged in the course of the run; its DP $s_1$ varies as PLT~2 is emerging and then decreases slowly. A PLT~2 is detected early after the first sink formation. Its slope $q_2$ rapidly steepens to $-2$ and then stabilizes at this value. In contrast, the DP $s_2$ initially swings within 2 orders of magnitude and seems to tend toward a constant value at $t\sim t_{\rm end}$. 

In the case of pure infall, a single sink particle forms (cf. Table \ref{table_W20_runs}, column 5) in the center of the clump. The spherical symmetry is preserved thereafter and it is instructive to compare the evolution of the radial density profile $\rho(l)$ (where $l$ is a given radius) to that of the \rhopdf{} (Appendix \ref{Density_profile_runs}). In case of a power-law density profile $\rho\propto l^{-p}$, its exponent is $p=-3/q$ \citep[see][and the references therein]{Donkov_Veltchev_Klessen_17}. Hence the obtained PLT slopes $q_2\simeq-2$ and $q_1\simeq-1.3$ in the pure-infall case correspond to $p_2=3/2$ and $p_1\simeq2.3$, respectively. For reference, in the classical Larson-Penston (LP) models of isothermal spherical collapse  \citep{Larson_69, Penston_69}, the structure of a protostellar object is characterised by profiles $p_{\,{\rm LP},\,2}=3/2$ (a core in free fall)  and $p_{\,{\rm LP},\,1}=2$ (its envelope where the infall motions are retarted due to thermal pressure support\footnote{However, \citet{Li_18} demonstrated that the profile $p=2$ is an intrinsic feature of scale-free gravitational collapse given that the central core (of very small size) has already collapsed and the rest of the clump has reached a quasi-steady state.}). The profile $p_1\simeq2.3$ in the outer zone around the sink particle, calculated directly (Fig. \ref{fig_profile_panel_figure}, right) or assessed from the slope $q_1$ of PLT~1 (Fig. \ref{fig_PLT_evolution_PI_density}, bottom), is steeper than the theoretical one. However, unlike the LP models, the gas in the simulations of W20 is non-isothermal, as illustrated in Fig.~\ref{fig_PI_density_temperature}, where we plot the density PDF together with the median gas temperature. Hence the support of thermal pressure against gravitational contraction is much stronger which leads to a flatter PLT~1. We point out that such an effect is not expected in local star-forming clumps wherein the temperatures are much lower due to the efficient cooling.

To assess the dynamics of the PDF evolution, it is appropriate to calculate $(t_{\rm end}-t_{\rm SF})$ in units of the free-fall time of the collapsing clump core. The latter is associated with the volume delineated by isodensity contour $\rho=\rho(\langle s_2 \rangle)$ where the parameters of PLT~2 are averaged in time: $\langle s_2 \rangle = 8 \ln(10)$ and $\langle q_2 \rangle = -2$ (see Fig. \ref{fig_PLT_evolution_PI_density}). With the mean density within this contour  
\[ \overline{\rho/\rho_0}=\frac{\lint_{\langle s_2 \rangle}^\infty \exp(s)p(s)\,\D s}{\lint_{\langle s_2 \rangle}^\infty p(s)\,\D s} = \frac{\langle q_2 \rangle}{\langle q_2 \rangle + 1} \exp(\langle s_2 \rangle)=2\times10^8~,   \]
the free-fall time is:
\[ \tau_{\rm ff} [\rho(\langle s_2 \rangle)]=(3\uppi/32 G \bar{\rho})^{1/2}\simeq 0.06~{\rm kyr}~. \]

Using the latter as an additional unit in Fig.~\ref{fig_PLT_evolution_PI_density} (top $x$-axis), one sees that both detected PLTs preserve nearly constant values of their parameters within 10 free-fall times of the collapsing clump core, i.e. the two PLTs reflect a quasi-stationary regime at this stage of the contracting clump evolution.

\begin{figure} 
% % \hspace{3.5em}
\begin{center}
\includegraphics[width=85mm]{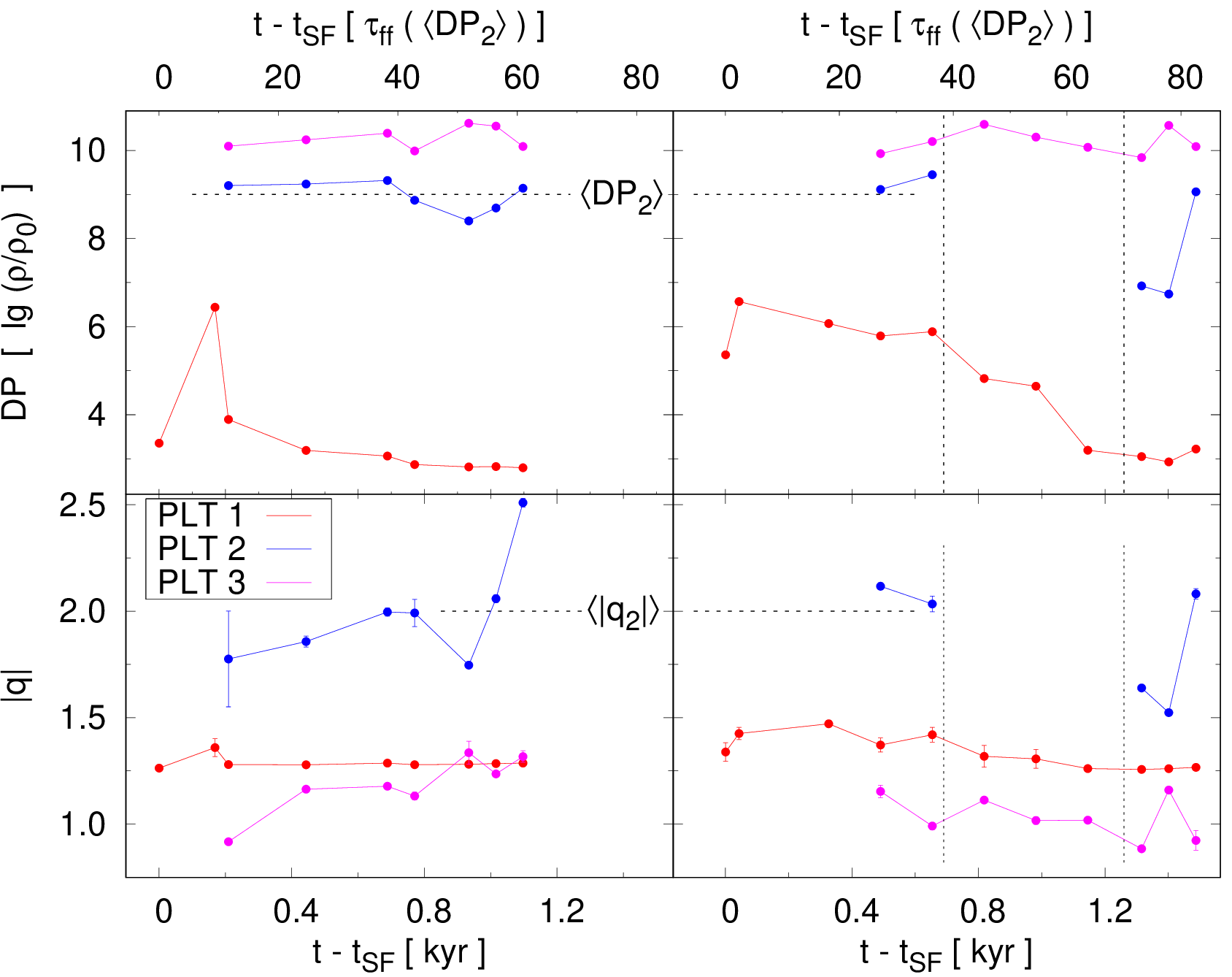}\\
\caption{The same as Fig. \ref{fig_PLT_evolution_PI_density}, but for two rotation-only runs: $\beta001-4$ (left) and $\beta010-2$ (right). Vertical dashed lines denote time spans when the PLT~1 and the PLT~2 merge into a single PLT. Other notations are the same like in Fig. \ref{fig_PLT_evolution_PI_density}.}
\label{fig_PLT_evolution_RO_density}
\end{center}
\end{figure}

\subsubsection{Local collapses with initial rotational support}
\label{rho-pdfs_rotational_support}

In the runs with initial rotational support, dozens of sink particles form, due to emergence of an unstable accretion disk and its subsequent fragmentation \citep[see also][]{Clark_ea_11, Prole_ea_23}. Their evolution in space, time and number is complex which leads to a permanent, dynamical re-distribution of material in the simulated contracting clump. This is clearly noticeable in the evolution of the PLTs shown in Fig. \ref{fig_PLT_evolution_RO_density}. In all  realizations with $\beta=0.01$ and $\beta=0.10$ we study, the slope of PLT~1 is the only PLT parameter whose value remains relatively stable in time. It varies within a narrow range around the mean value obtained in the pure-infall case ($-1.25 \gtrsim q_1 \gtrsim -1.5$) and appears not to be significantly affected by the location of the DP $s_1$ and/or the evolution of other detected PLTs.

Both PLT~2 and PLT~3 appear early after $t=t_{\rm SF}$ in the realizations with $\beta=0.01$ and are detectable up to $t\lesssim t_{\rm end}$ (Fig. \ref{fig_PLT_evolution_RO_density}, left). They are of similar length, with PLT~3 being much shallower. Although $s_2$ and $s_3$ vary permanently, the minimal length of PLT~2 (about 1 decade) is being preserved. PLT~2 steepens slowly in the course of the run, slower than in the pure-infall case (cf. Fig. \ref{fig_PLT_evolution_PI_density}, bottom), but eventually reaches larger slopes, still comparable to $-2$ -- i.e., the result from the pure infall is roughly reproduced. The slope of PLT~3 varies around $-1$, with deviations up to $0.3$. In terms of scale, PLT~2 and PLT~3 respectively correspond to effective sizes of dozens to a few hundred AU (vicinity/envelope of the emerging protocluster) and $\sim 10-20$~AU; see Fig. \ref{fig_Size-slope_relation}. The latter scales are typical for thin accretion disks which can form from 2D turbulence -- allegedly one of the mechanisms for transfer of angular momentum outwards \citep[e.g.,][]{Dubrulle_Valdettaro_92, Klahr_Bodenheimer_03}. In agreement with such interpretation, \citet{Donkov_ea_24} showed that a PLT with $q\sim -1$ should emerge at the high-density end of the \rhopdf{} in self-gravitating, polytropic turbulent clouds -- this PLT corresponds to a thin, rotating accretion disk in the close vicinity of a protostellar core.

In the runs with stronger initial rotational support ($\beta=0.10$), a PLT~3 emerges and behaves in a similar way as for $\beta=0.01$. However, here PLT~2 is very unstable and merges with PLT~1 at some points in time (Fig. \ref{fig_PLT_evolution_RO_density}, right). The latter phenomenon can be explained from inspection of the (column-)density and temperature maps at late evolutionary stages of the collapsing clump (Fig. \ref{fig_RO_runs_maps} in the Appendix). In the runs with $\beta=0.10$ the emerging protocluster is more dispersed in space and consists of several subgroups with total effective size of the PLT~2 regime -- their interaction causes complex dynamics of gas flows and perturbations of the density field.

\subsubsection{Local collapses with initial turbulent support}
\label{rho-pdfs_turbulent_support}

\begin{figure} 
% % \hspace{3.5em}
\begin{center}
\includegraphics[width=84mm]{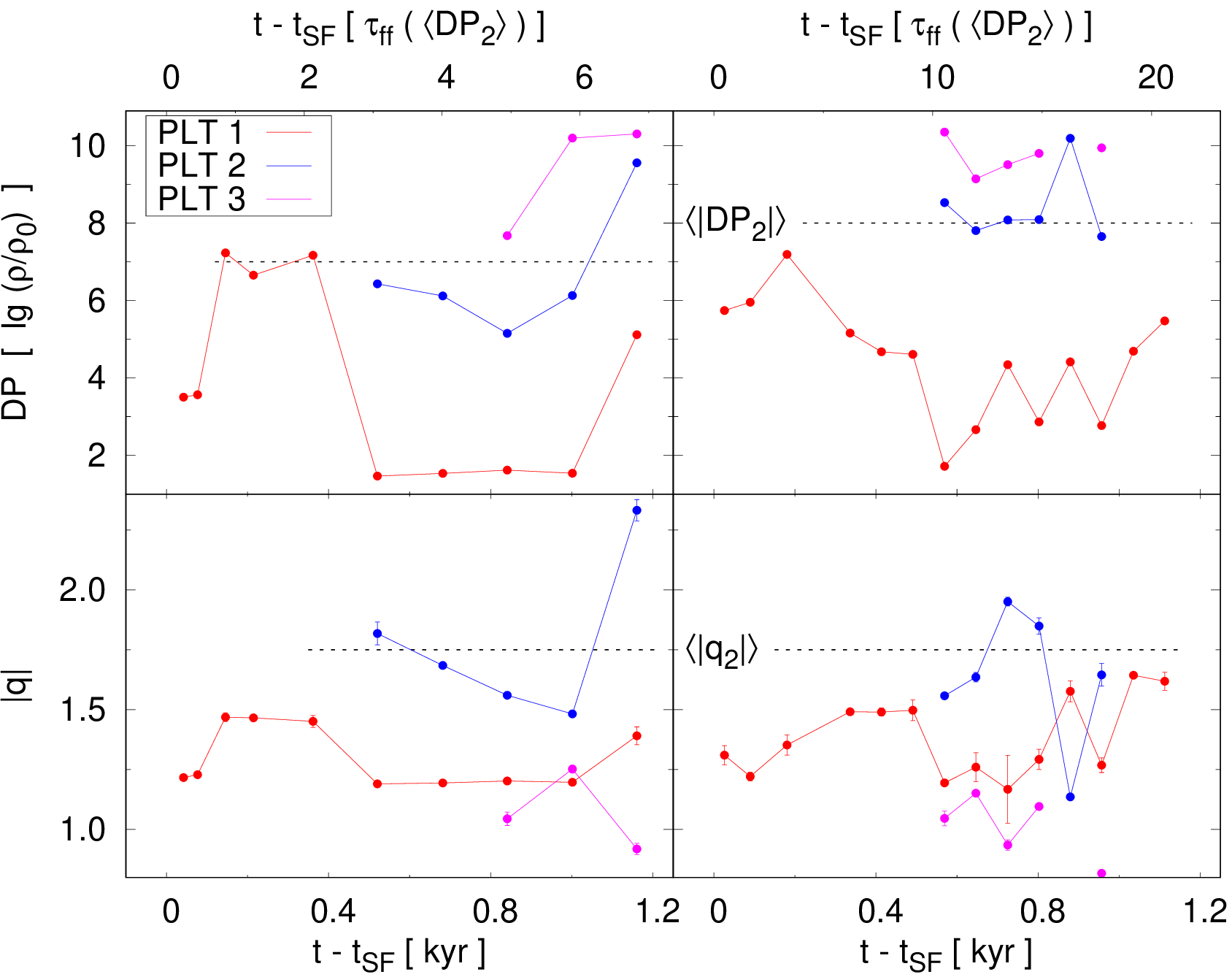}\\
\caption{Evolution of PLTs of the \rhopdf{} in runs with initial turbulence: $\alpha025-1$ (left) and $\alpha025\beta010-1$ (right). The notations are the same like in Fig. \ref{fig_PLT_evolution_RO_density}.}
\label{fig_PLT_evolution_TO+TR_density}
\end{center}
\end{figure}

Runs with $\alpha>0$ and $\beta=0$ are also characterized by some non-zero initial angular momentum (see W20). For comparison with the results of the RO runs, we select one TO and one TR realization with strong turbulent support, i.e. with $\alpha=0.25$ which corresponds to virial parameter of $\lesssim 1$, given the initial density profile \citep{McKee_Zweibel_92}. In both cases, $t_{\rm SF}$ and $N_{\rm sink}$ are similar to those of the runs with initial rotational support, displaying a bit larger star-formation efficiency (Table \ref{table_W20_runs}), whereas the \rhopdf{} evolution is more complex. In the TO run $\alpha025-1$, PLT~2 emerges initially with slope $|q_2|\lesssim 2$ -- like the one from the PI runs but significantly later in time (Fig. \ref{fig_PLT_evolution_TO+TR_density}, left). Its further evolution is seemingly intertwined with the development of a PLT~3 which becomes detectable at times close to the end of the run. At $t=t_{\rm end}$, the \rhopdf{} consists of 3 PLTs, with parameters quite similar to those from the RO runs.

Figure \ref{fig_PLT_evolution_TO+TR_density}, right, exemplifies the combined effect of strong initial rotational and turbulent support. The emergence of multiple PLTs occurs at about the same time as in the $\beta010$ runs (cf. Fig. \ref{fig_PLT_evolution_RO_density}, right). However, {\it both} PLT~2 and PLT~3 are unstable in time, merge and eventually disappear at $t\sim t_{\rm end}$. Such a picture is expected in view of intermittent turbulence. The general, the conclusions are that the inclusion of turbulent support slows down the development of multiple PLTs and that the presence/absence of the latter cannot be used as strong indicators of physical processes in turbulent, collapsing regions.

Nevertheless, PLT~3 of slope $q_3\sim-1$ appears in all runs with rotational and/or turbulent support. The spatial scale and the density range of this regime corresponds to the sizes of the formed dense accretion disks. Obviously the PLT~3 range is indicative of the role of centrifugal forces in the contracting clump, as suggested originally by \citet{Kritsuk_Norman_Wagner_11}.

\subsection{Column-density PDFs}
\label{N-PDFs_evolution}

\begin{figure} 
% % \hspace{3.5em}
\begin{center}
\includegraphics[width=84mm]{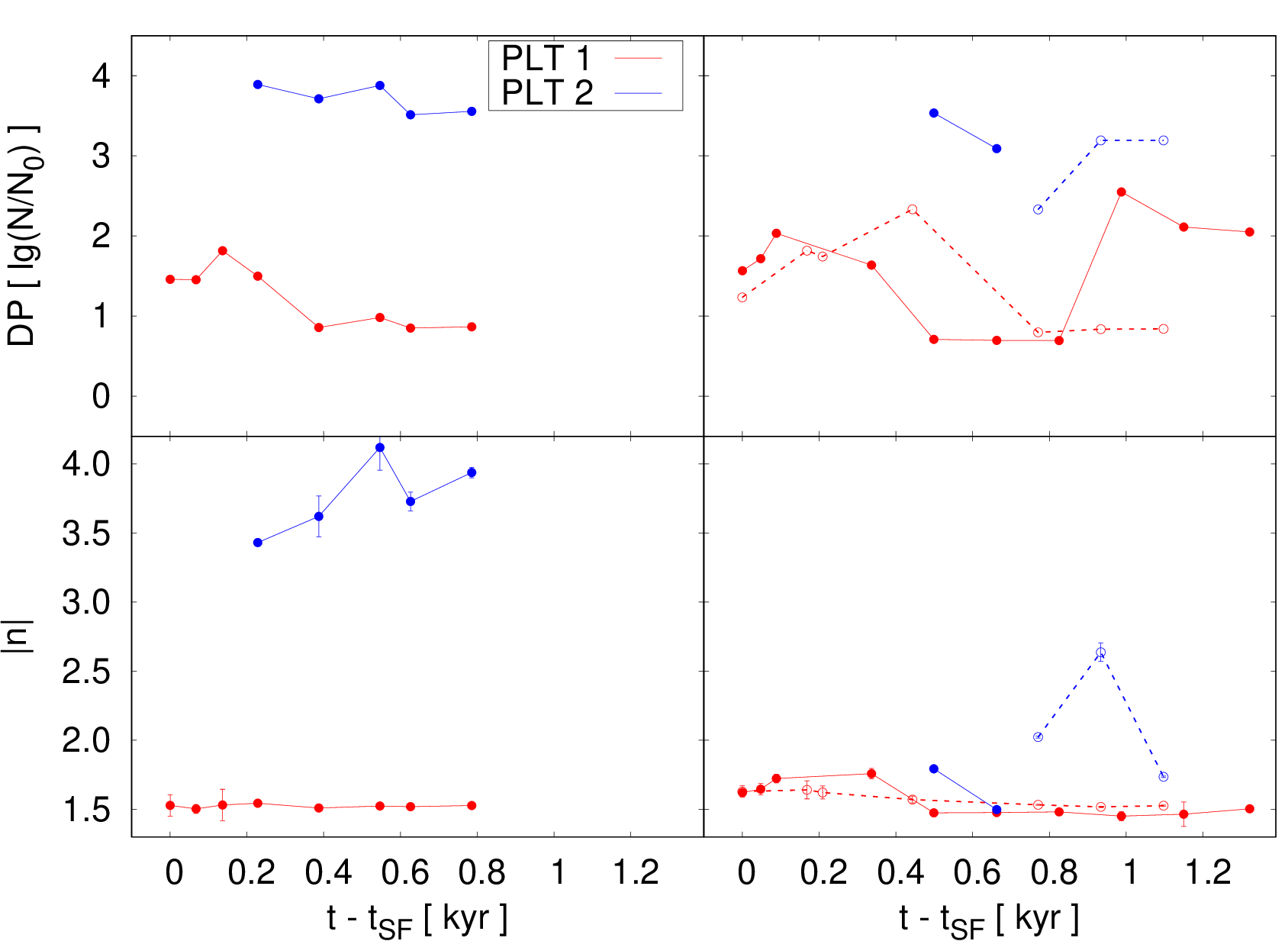}\\
\caption{Evolution of PLTs of the \Npdf{} (projection along the $Z$ axis) in the pure-infall case (left) and in cases with initial rotational support (right): $\beta=0.01$ (dashed, open symbols) and $\beta=0.10$ (solid, filled symbols). A single realization is chosen for each case.}
\label{fig_NPLT_evolution}
\end{center}
\end{figure}

The \Npdf{}s in the simulations of W20 also display multiple PLTs. The evolution of the PLT parameters derived from column-density maps obtained from the projection along the rotational $Z$-axis is illustrated in Fig. \ref{fig_NPLT_evolution}. The essential difference with the \rhopdf{} investigation is that a PLT~3 is not resolved at all. PLT~1 is apparently a permanent feature in the pure-infall case as well in those with initial rotational support. Its slope varies in the range $-1.75\lesssim n_1 \lesssim -1.45$. PLT~2 is detectable shortly after $t=t_{\rm SF}$ in the pure-infall case, with slope which increases towards $-4$ by the end of the run. In the runs with rotational support, PLT~2 is much flatter; for $\beta=0.10$ it merges with PLT~1 -- shortly after its emergence and analogously to its density counterpart (Fig. \ref{fig_NPLT_evolution}, top right).

\begin{figure}
% \hspace{3.5em}
\begin{center}
 \includegraphics[width=85mm]{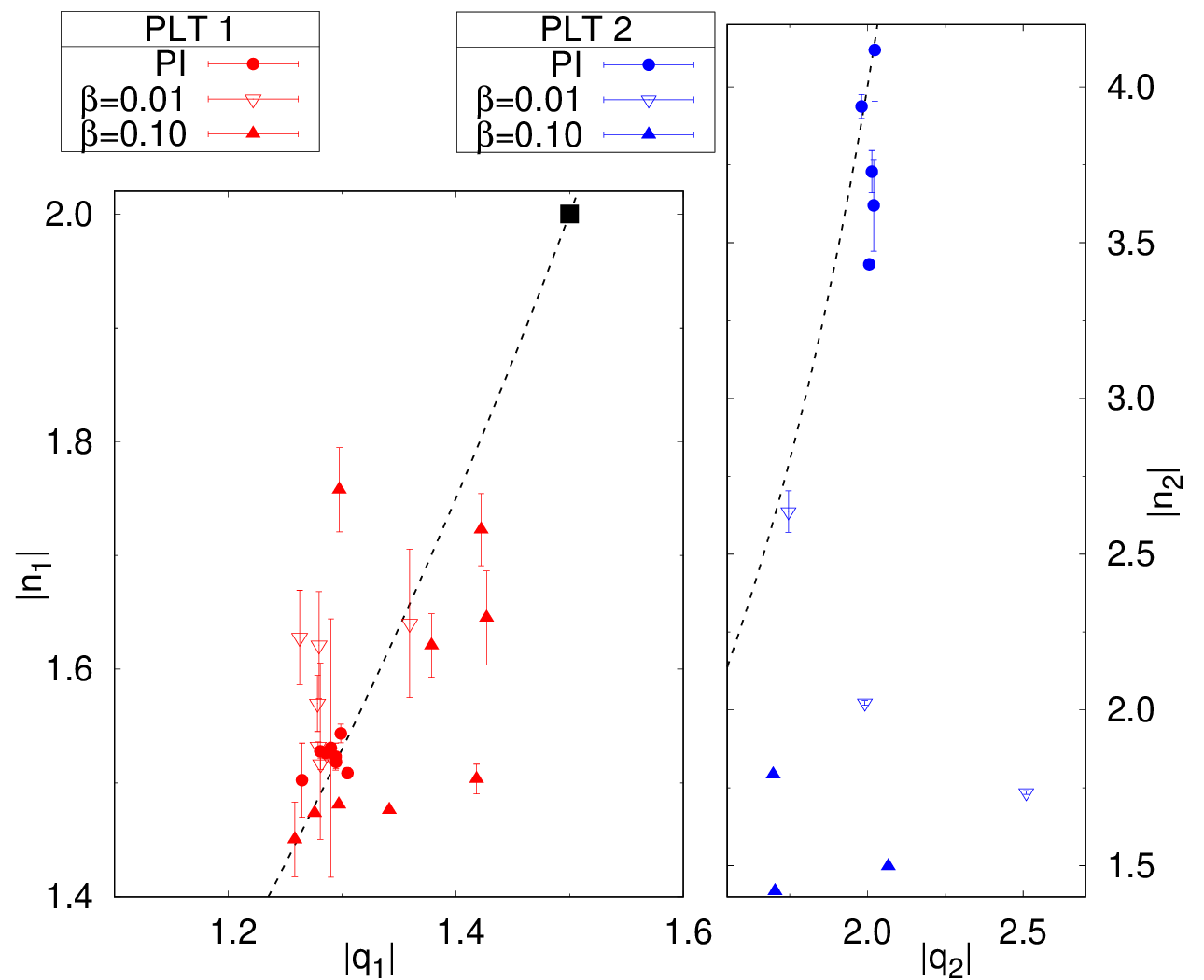}
\caption{Correspondence between the slopes of PLT~1 (left) and PLT~2 (right) from \rhopdf{}s and their \Npdf{} counterparts from maps averaged over the three main directions. The expected relation from equation \ref{eq_q-n_relation} is plotted for reference (dashed). The black square in the left panel shows the expectation from the LP model of collapsing core; see text.}
 \label{fig_Slopes_PLTs_relationship}
\end{center}
\end{figure}

An interesting issue in the PDF analysis is the relationship between the PLT slopes $q$ and $n$ in the $\rho$- and \Npdf, respectively. It has been demonstrated in a number of works \citep[see, e.g.][Sect. 5.2, and the references therein]{Donkov_Veltchev_Klessen_17} that, if the cloud structure can be described by a power-law density profile (i.e. in case of point symmetry in regard to the clump centre), those PLT paratemers should relate as: 
\begin{equation}
\label{eq_q-n_relation}
 n=2q/(3+q)~.
\end{equation}
Since the clumps in this study are self-gravitating and contracting, a power-law density profile is expected, at least in the inner regions, while the presence of rotating disk-like structures should break the spherical symmetry. In Fig. \ref{fig_Slopes_PLTs_relationship} we juxtapose the derived slopes $n_1$ (left) and $n_2$ (right) plotted in Fig. \ref{fig_NPLT_evolution} with their counterparts from the \rhopdf{} analysis and plot the relation from equation (\ref{eq_q-n_relation}) for comparison. In the runs with pure infall, the correlation between $q_1$ and $n_1$ is consistent with the theoretical expectation, with an apparent systematic shift with respect to the slope $q_{\,{\rm LP},\,1}=-3/2$ ($n_{\,{\rm LP},\,1}=-2$) corresponding to the asymptotic density profile $p=2$ in the envelope of collapsing core in free fall \citep{Larson_69, Penston_69}. We attribute this phenomenon to the non-isothermal thermodynamics in the W20 simulations commented on in Sect. \ref{Pure-infall_rho_pdfs} and leading to shallower slopes $q_1$ and, hence, $n_1$. The results from runs with low rotational support display similar behaviour with some larger scatter while those from the run with $\beta=0.10$ deviate significantly from relation (\ref{eq_q-n_relation}) due to the merger of PLT~2 with the PLT~1. The diagram $q_2$ vs. $n_2$ (Fig. \ref{fig_Slopes_PLTs_relationship}, right) displays an even clearer segregation between the pure-infall case and the ones with initial rotational support. The slopes of the second PLTs in the pure-infall case still relate in a good agreement with the expectation from the LP collapse model unlike those (if detected at all) in the runs $\beta=0.01$ and $\beta=0.10$. The latter finding indicates that in case of rotational support the spherical symmetry is broken at the spatial scales which correspond to PLT~2.

\section{Discussion}
\label{Discussion}

\subsection{Comparison with other works on PLTs}
\label{Discussion_PLTs_works}
In general, our result on the \rhopdf{} in the pure-infall case (Sect. \ref{Pure-infall_rho_pdfs}) reproduces the solution from the classical LP protostellar collapse model \citep{Larson_69, Penston_69}: a protostar formed within a core ($q_2\simeq-2$; $p_2\simeq 3/2$) with its outer envelope ($q_1\simeq-1.3$; $p_1\simeq 2.3$), where $p_1>p_{\,{\rm LP,}\,1}=2$ due to the non-isothermal thermodynamics and, hence, stronger pressure support. This picture is not surprising in view of the setup of the used W20 simulations: collapsing BE spheres put in a large, homogeneous gas reservoir. The two protostellar `core + envelope' regimes (corresponding to PLT~2 and PLT~1, respectively) are recognizable even in the runs with initial rotational and/or turbulent support in which dozens of protostars form, with a complex and dynamical distribution in space.   

Accretion on the protostar(s) is far from steady state and its total rate varies significantly in all studied runs (see Sect. 4.3 in W20). This clearly affects the structure and evolution of the PDF. Another and more appropriate model for comparison seems to be the one of \citet{Murray_Chang_15} which is aimed to describe self-gravitating clumps wherein massive star formation takes place. Taking into account the driving of turbulence in the compressed accretion flows, those authors find an `attractor solution' at small scales where the gravity of the protostar dominates: a time-independent density profile with $p=3/2$, irrespectively of the type of support. At larger scales, the profile exponent should depend on position: $1.6\lesssim p(\rho) \lesssim 1.8$. This picture is supported by simulations of \citet{Murray_ea_17}. In terms of \rhopdf{} (see their Fig. 10), they found a PLT with slope $\sim-1.7$ which seems to be a combination of PLT~1 and a steeper PLT~2 and another, very shallow one at the high-density end, with slope $>-1$. This general \rhopdf{} shape becomes clearer when only a sphere of 1 pc radius around the star particle is considered and is approximately consistent with our results.

The effects of turbulence on the \rhopdf{} evolution in isothermal gravoturbulent fluids were recently studied by \citet{Khullar_ea_21}, based on simulations with systematic variation of the Mach number and the virial parameter. The main PDF part is initially lognormal due to the driven turbulence. Later on two PLTs develop which reflects the increasing role of gravity and rotation -- the DP of the first one marks the transition from unbound to bound gas and that of the second one delineates disk structures in the collapsing region. In terms of slope, the two PLTs detected in \citet{Khullar_ea_21} remain relatively stable in the course of runs with stronger self-gravity and generally correspond to PLT~2 and PLT~3 extracted in this study: $q_2\sim -2$ and $q_3\sim -1$. We note, however, that the density span of PLT~2 in their work is much larger which can be attributed to the strong turbulent support against collapse.

\subsection{Comparison with observations}
\label{PLTs_comparison_with_Herschel}
Although many recent works discuss the features of \Npdf{}s derived from observations of star-forming regions, only a few are dedicated to the multiple-PLT phenomenon. The study of \citet{Schneider_ea_22} based on {\it Herschel} maps is the only one (to our knowledge) which provides a large statistics -- those authors found evidence for double PLTs in \Npdf{}s of 28 Galactic regions of various mass, size and star-forming activity: from giant molecular clouds with signatures of high-mass formation to molecular clouds wherein low-mass stars form (LM regions) and diffuse regions without star formation. The objects from their sample which might be relevant to compare with the simulated contracting clumps from this paper in terms of mass and size are the LM regions (cf. Table 1 in \citealt{Schneider_ea_22}). The latter are also well resolved at the scales of starless/prestellar cores ($<0.1$~pc). On the other hand, in view of the very different chemistry, temperature regime and timescales of star formation, the comparison presented below should be considered with caution.

We apply the adapted \bPLFIT{} technique to the \Npdf{}s of 10 LM regions to extract PLTs. In many cases, the derived PLT parameters differ significantly from those of \citet{Schneider_ea_22} who fitted the \Npdf{}s by use of sets of models which combine lognormal and power-law functions and determined the best fitting model through the Bayesian information criterion. This is not surprising in view of the essential difference between the methods applied; see Appendix \ref{BIC_and_bPlfit methods} for an illustration and comments.

\begin{figure}
% \hspace{3.5em}
\begin{center}
 \includegraphics[width=85mm]{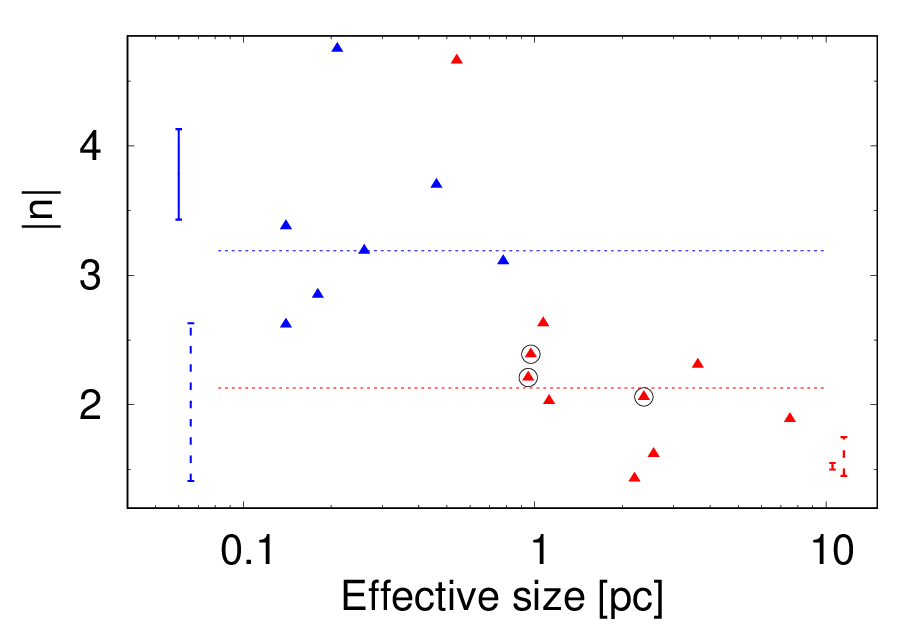}
\caption{Slopes of PLT~1 (red) and PLT~2 (blue) of \Npdf{}s, derived from {\it Herschel} maps of low-mass star-forming regions \citep{Schneider_ea_22}, plotted against the spatial scale. Open circles denote detections of a single PLT; median values $|n_{1,\,{\rm med}}|$ and $|n_{2, \,{\rm med}}|$ are shown with dotted lines. The ranges of slope variation in the pure-infall case (solid line) and in those with rotational support (dashed line) from this work (Sect. \ref{N-PDFs_evolution}) are given for comparison.}
 \label{fig_Double_PLTs_in_SFR}
\end{center}
\end{figure}

In Fig. \ref{fig_Double_PLTs_in_SFR} the derived slopes are plotted vs. the spatial scales defined by the given PLT range of column density. The effective sizes of the scales are calculated analogously to those shown in Fig. \ref{fig_Size-slope_relation}: $l_{\rm eff}=(S({\rm PLT}_i)/\uppi)^{1/2}$ where $S({\rm PLT}_i)$ are the areas\footnote{The used distances to the regions are given in Table 1 in \citet{Schneider_ea_22}.} enclosed by the isocontours corresponding to the DPs of PLT 1 and PLT 2 ($1\le i \le2$). Single PLTs of similar slope are detected in three regions (open circles). The PLT~1 and PLT~2 regimes seem to be clearly separated in terms of spatial scale: the former are comparable to the size of the whole star-forming region whereas the steeper\footnote{With the exception of Pipe region where a very steep PLT~1 is derived.} PLT~2 is associated with clumps of typical size a few  tenths of pc. The median slope values of PLT~1 $|n_{1,\,{\rm med}}|=-2.13$ (red dotted line) and PLT~2 $|n_{2,\,{\rm med}}|=-3.2$ (blue dotted line) are much closer than those found from our PI runs but the scatter is too large to allow for a clear conclusion.

To sum up, the shape of the \Npdf{}s from this work is similar to the well resolved high-density part of the ones in all but one LM star-forming regions studied by \citet{Schneider_ea_22}.

\section{Summary}
\label{Summary}

This work presents a numerical study of the evolution of power-law tails (PLTs) in (column-)density probability distribution functions (\rhopdf{} and \Npdf{}). We make use of high-resolution simulations of contracting star-forming clumps in primordial gas, with varied initial (rotational, turbulent) support against gravity \citep{Wollenberg_ea_20}, and of the adapted \bPLFIT{} method for PLT extraction \citep{Veltchev_ea_19, Marinkova_ea_21} which assesses the assumed PLT without any assumptions about the possible functional shape of the rest of the PDF. The clumps are strongly dominated by self-gravity and we are interested in those density ranges where it plays the major role. Therefore the main features of the collapse phase should not differ substantially from those in the present-day star-forming regions.

The primary results can be summarized as follows:
\begin{itemize}
 \item In all considered runs (pure infall, initial rotational support and/or initial turbulent support) multiple PLTs emerge shortly after the formation of the first protostar. The first PLT (PLT~1) in \rhopdf{} is a stable feature with slope $q_1\simeq -1.3$ which corresponds to the outer envelope of the protostellar object (with density profile $\rho\propto l^{-2}$) in the classical Larson-Penston (LP) protostellar collapse model \citep{Larson_69, Penston_69} if the assumption of spherical symmetry is valid. The shallower slope is explained with the non-isothermal thermodynamics which provides additional support against collapse.
 \item The second PLT (PLT~2) in \rhopdf{} is a stable feature in the pure-infall runs (a single protostar formed) but fluctuates significantly in the runs with initial motions (rotation/turbulence) as dozens of protostars form. Its mean slope $\langle q_2\rangle\simeq -2$ corresponds to a density profile $\rho\propto l^{-3/2}$ which describes a core in free fall, in the LP collapse model, or an attractor solution at scales with domination of protostar's gravity, in the model of self-gravitating clumps of \citet{Murray_Chang_15}. 
 \item PLT~1 and PLT~2 in the \Npdf{}s from the pure-infall runs and those with initial rotational support are generally consistent with the estimates obtained through the adapted \bPLFIT{} method from {\it Herschel} data for a number of Galactic low-mass star-forming regions \citep{Schneider_ea_22}. 
 \item In the runs with initial rotation and/or turbulence a third PLT (PLT~3) in \rhopdf{}s appears simultaneously with or after the emergence of PLT~2. It is very shallow and is associated with formation of protostellar accretion disks. Its mean slope $\langle q_3\rangle\simeq -1$ agrees with the results from numerical studies of supersonic, isothermal, self-gravitating turbulence \citep{Kritsuk_Norman_Wagner_11, Khullar_ea_21}.
\end{itemize}

\section*{Acknowledgement} 
We thank the anonymous referee whose critical comments helped to improve the paper. The authors are grateful to K. Wollenberg for the numerical data placed at our disposal and to N. Schneider for the \textit{Herschel} maps of Galactic star-forming regions used solely to derive the column-density distributions.

T.V. and S.D. acknowledge support by the Deutsche Forschungsgemeinschaft (DFG) under grant KL 1358/20-3 and additional funding from the Ministry of Education and Science of the Republic of Bulgaria, National RI Roadmap Project DO1-176/29.07.2022. P.G. acknowledges funding from the European Research Council via the ERC Synergy Grant ``ECOGAL'' (grant 855130). 
T.V. acknowledges the access to the Nestum cluster \@ HPC Laboratory, Research and Development and Innovation Consortium, Sofia Tech Park, Bulgaria.
R.S.K.\ also acknowledges financial support  from the ERC via the ERC Synergy Grant ``ECOGAL'' (project ID 855130),  from the Heidelberg Cluster of Excellence (EXC 2181 - 390900948) ``STRUCTURES'', funded by the German Excellence Strategy, and from the German Ministry for Economic Affairs and Climate Action in project ``MAINN'' (funding ID 50OO2206). 
The authors gratefully acknowledge the data storage service SDS@hd supported by the Ministry of Science, Research and the Arts Baden-Württemberg (MWK) and the German Research Foundation (DFG) through grant INST 35/1503-1 FUGG, and they thanks for computing resources provided in bwHPC through grant INST 35/1134-1 FUGG. 

\section*{Data availability}
No new data were generated or analysed in support of this research.

\label{lastpage}
\appendix
\newpage

\section{Density profile from the pure-infall runs}
\label{Density_profile_runs}

\begin{figure*}
% \hspace{3.5em}
\begin{center}
 \includegraphics[width=0.85\textwidth]{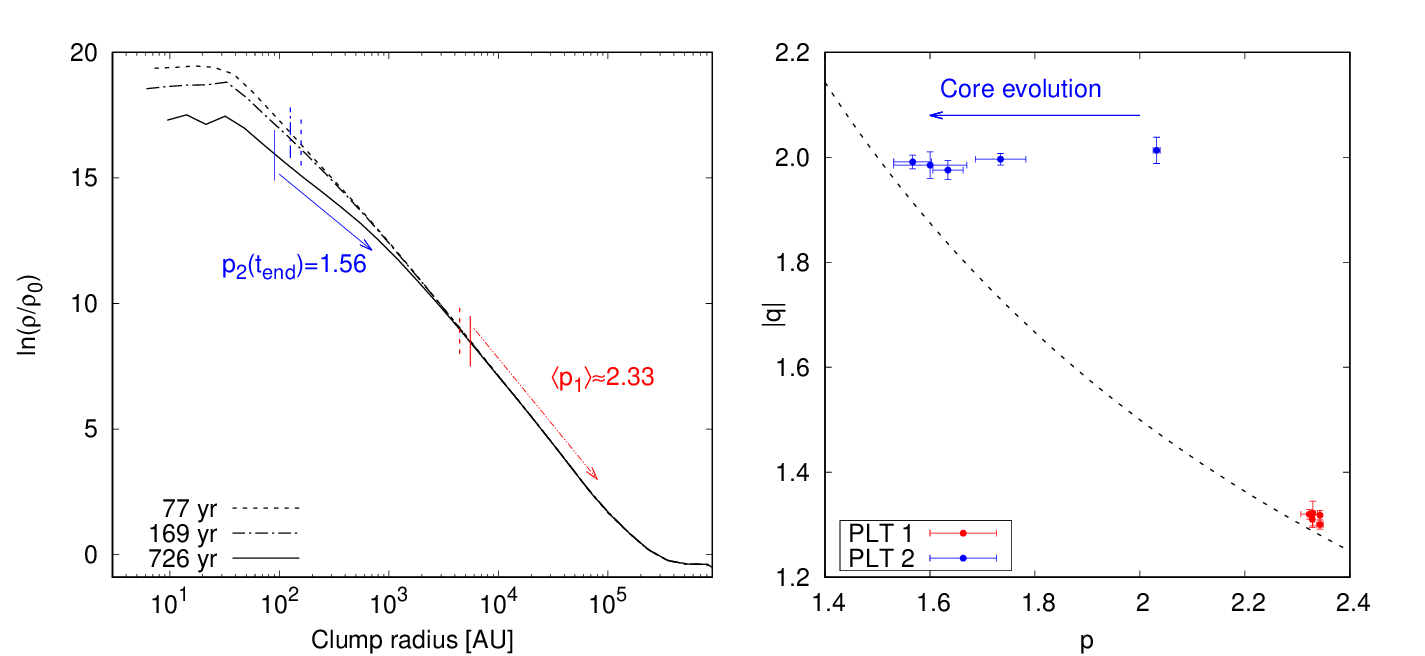}
\caption{Evolution of the density profile from a pure-infall run (left) and correspondence between its PL exponents $p$ and the PLT slopes $q$ of the \rhopdf{} (right). Vertical marks in the left panel denote the deviation points of the deviation points of the power-law parts for the selected evolutionary times (in units $t-t_{\rm SF}$). Dashed line in the right is the expected relation $q=-3/p$; see text.}
 \label{fig_profile_panel_figure}
\end{center}
\end{figure*}

In the runs with pure-infall motions the spherical symmetry is preserved up to $t=t_{\rm end}$. This allows to study the evolution of the density profile $\rho=\rho(l)$ within the contracting star-forming clump. We applied the adapted \bPLFIT{} method (Sect. \ref{Method}) to extract the power-law (PL) parts of the density profile $\rho\propto l^{-p}$  (Fig. \ref{fig_profile_panel_figure}, left). At times $t\simeq t_{\rm SF}$, a PL part of the profile forms in the outer shells of the clump and preserves its slope $p_1$ and deviation point in the course of the further contraction. A second PL part evolves in the near vicinity of the small collapsing (quasi-homogeneous) core containing the sink particle; its slope $p_2$ gets shallower and approaches $1.5$ at the end of the run. This picture is in general agreement with the Larson-Penston (LP) models of isothermal spherical collapse \citep{Larson_69, Penston_69} in which $p_{{\rm LP},\,2}=3/2$ is the profile of the core in free fall and $p_{{\rm LP},\,1}=2$ reflects the retarded motions in its envelope. The steeper slope $p_1$ obtained in this study has to be attributed to the non-isothermality of the gas (see Sect. \ref{Pure-infall_rho_pdfs}).

The right panel of Fig. \ref{fig_profile_panel_figure} juxtaposes the PL exponents of the density profile with the extracted PLTs of the \rhopdf. As shown by different authors (see Sect. 5.2 in \citealt{Donkov_Veltchev_Klessen_17} and the references therein), the correspondence between these quantities should be $q=-3/p$ in case of spherical symmetry. The obtained values of $p_1$ are very close to the theoretical expectation while $p_2$ approaches it with the rapid evolution of the collapsing core.

\section{Late evolution of clumps with rotational support}
\label{Maps_RO_runs}

\begin{figure*} 
\begin{minipage}{\textwidth}
\includegraphics[width=\textwidth]{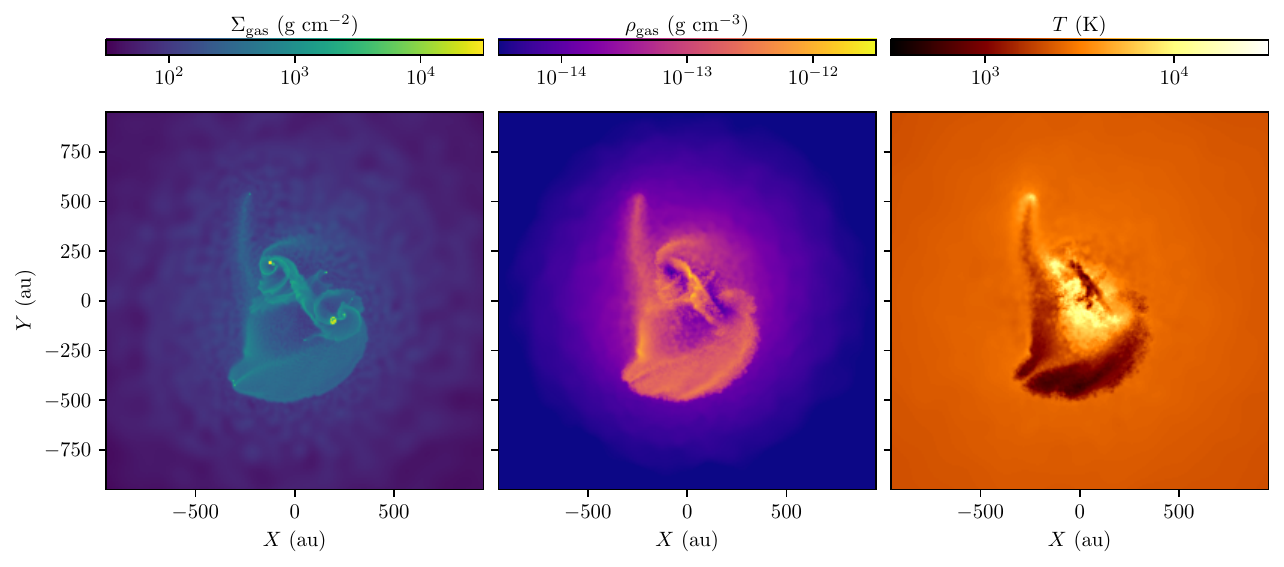}
    \includegraphics[width=\textwidth]{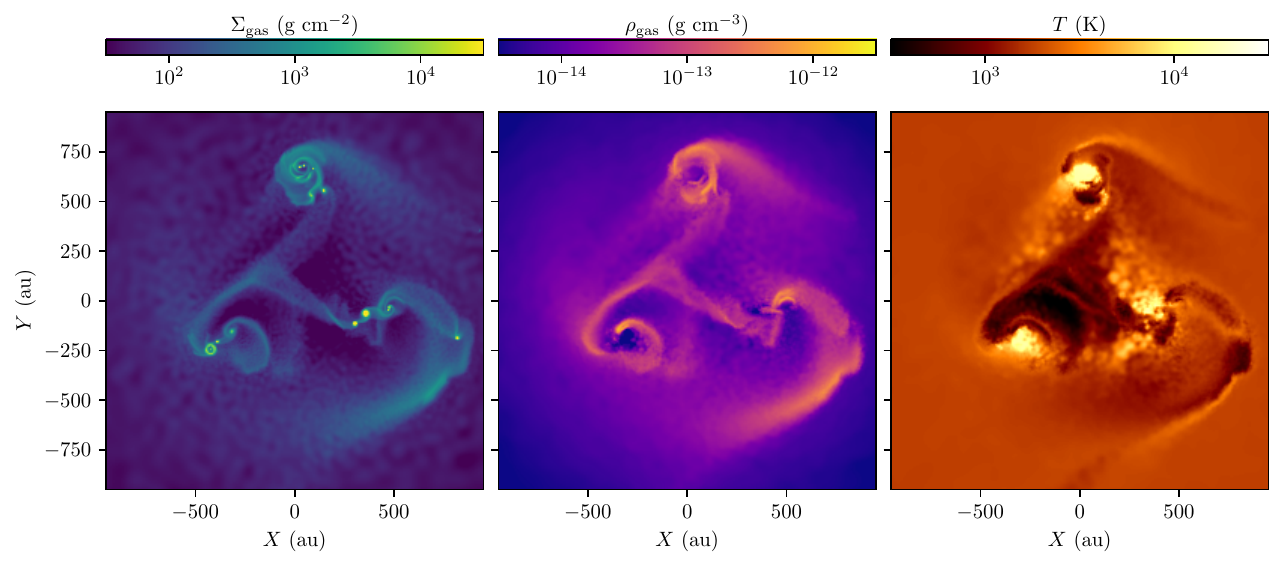}
\caption{Snapshot of column-density (left), density (center) and temperature (right) fields in the central clump region, at the end of the W20 runs with initial rotational support: $\beta001-4$ (top) and $\beta010-4$ (bottom). The slice of the density field is taken at the plane $z=0$.}
\label{fig_RO_runs_maps}
\end{minipage}
\end{figure*}

Dozens of protostars have formed at the final stage of the W20 runs with initial rotational support (cf. Table \ref{table_W20_runs}). They populate the central zone of the collapsing clump displayed in Fig. \ref{fig_RO_runs_maps}. As seen from the temperature map (right panels), the star-formation region with $T\sim 10^4$~K in the $b=0.01$ case is compact and centered at the clump core whereas in the $b=0.10$ case it consists of three distinct subregions. The (column-)density maps in the left and middle panels illustrate the gas dynamics. The PLT~3 regime in the \rhopdf{}s corresponds to densities $\gtrsim 10^{-10}$\gcc~in the very close vicinities of the protostars while the PLT~2 densities are typical for infalling flows towards them. Those flows are extended and spiral-like in the $b=0.10$ case (bottom center); their fluctuations at the final evolutionary stage cause the instability and/or the disappearance of the PLT~2 commented in Sect. \ref{rho-pdfs_rotational_support}.

\section{Multiple power-law tails from observational {\it N}-pdfs}
\label{BIC_and_bPlfit methods}

\begin{figure}
% \hspace{3.5em}
\begin{center}
 \includegraphics[width=85mm]{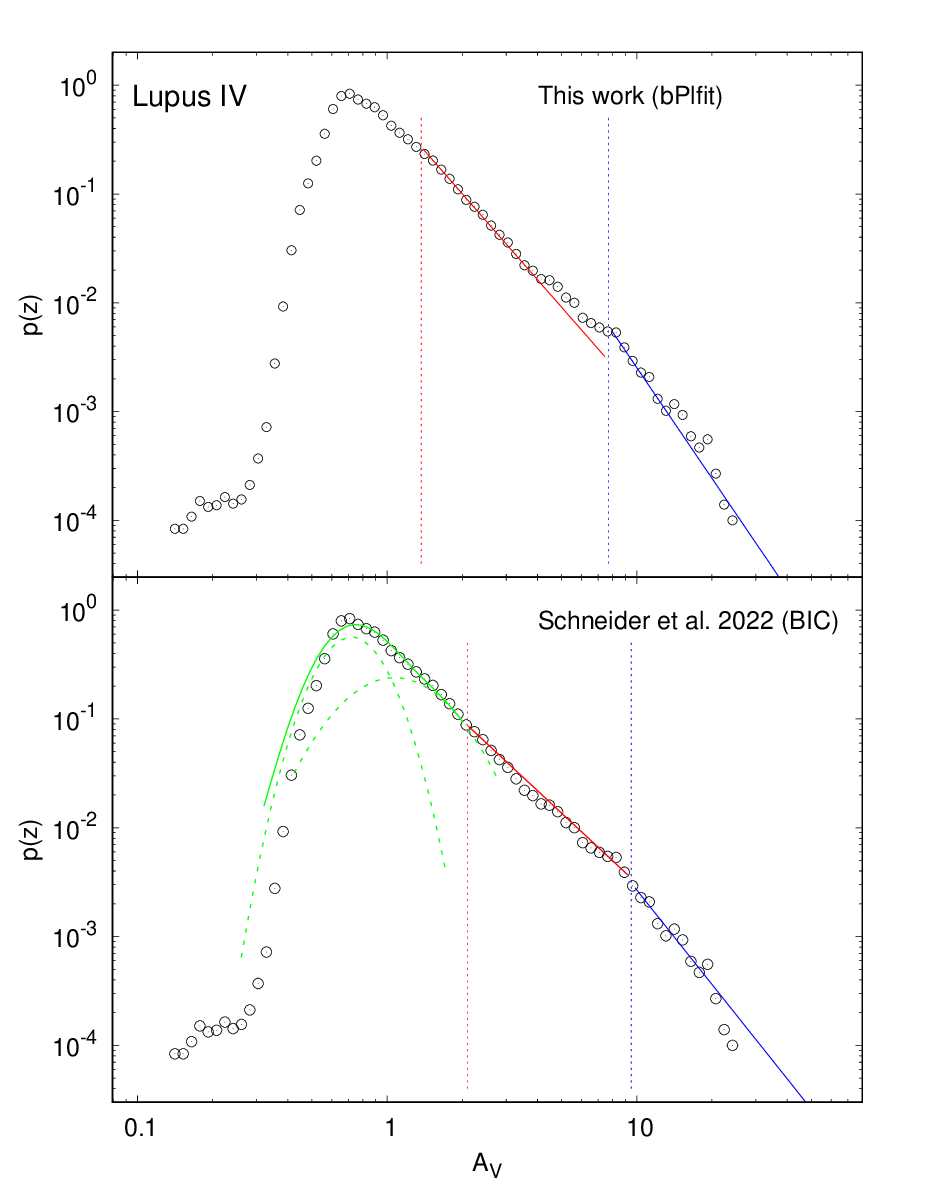}
\caption{Example of multiple PLT extraction through the methods used in \citet{Schneider_ea_22} (bottom) and in this work (top). The slopes of PLT~1 (red) and PLT~2 (blue) are shown; the corresponding DPs are marked with vertical dashed lines. The two lognormal functions (long-dashed) and their sum (solid) which fits the main \Npdf{} part in \citet{Schneider_ea_22} are plotted in green.}
 \label{fig_BIC_bPlfit_comparison}
\end{center}
\end{figure}

The extraction of multiple PLTs in \citet{Schneider_ea_22} and in the present work differ essentially in terms of presuppositions and technique. In the former paper, the basic presupposition is that the \Npdf{} is a combination of {\it at least} one lognormal and one power-law function (plus an additional PLT to account for `error slopes' at the low-density end); the best-fitting model is selected from a set of models through the Bayesian information criterion as the numbers of lognormals and PLTs are varied between one and two. An exemplary result from this approach is illustrated in Fig. \ref{fig_BIC_bPlfit_comparison}, bottom. Note that no gaps between the spans of PLT~1 and PLT~2 are allowed. In contrast, the adapted \bPLFIT{} method rests on the assumption of a single PLT only and is searching for a possible PLT~2 (see Sect. \ref{Method}). The column-density ranges of PLT~1 and PLT~2 can be separated due to \Npdf{} features (e.g., large bumps) which are poorly fitted by a power-law (Fig. \ref{fig_BIC_bPlfit_comparison}, top).\\
Both methods have their advantages in analysing the \Npdf{}s in star-forming regions. The approach of \citet{Schneider_ea_22} is based on the pertinent physical assumption that the main part of the PDF is shaped primarily by isothermal, non-gravitating, supersonic turbulent gas which is characterized by a lognormal density distribution (see references in Sect. \ref{Introduction}). In some regions, this main PDF part can be fitted by multiple lognormals which correspond to some characteristic spatial scales in zones of diffuse gas \citep{Stanchev_ea_15}. On the other hand, the adapted \bPLFIT{} method can capture PLTs which are separated in terms of density range due to -- for instance -- projection effects and/or the evolutionary stage of collapsing clouds.

\end{document}